\documentclass[fleqn,twoside,twocolumn,nofootinbib,showkeys]{revtex4} 
\usepackage[sec,nocpr]{ujp} 
\begin{document}
\title[Particle thermalization entropy, Unruh effect]
{PARTICLE THERMALIZATION ENTROPY AND UNRUH EFFECT}%
\author{M.V.~Teslyk}
\affiliation{Taras Shevchenko Kyiv University, Physics Department}
\address{Build. 1, 2, Glushkov Prosp., Kyiv 03680, Ukraine}
\email{machur@ukr.net}
\author{O.M.~Teslyk}
\affiliation{Taras Shevchenko Kyiv University, Physics Department}
\address{Build. 1, 2, Glushkov Prosp., Kyiv 03680, Ukraine}

\udk{539.12.01} \pacs{04.70.Dy, 12.38.Aw, 12.38.Mh, 25.75.-q}
\razd{\secix 1.2,11}

\autorcol{M.V.~Teslyk, O.M.~Teslyk}%

\setcounter{page}{1}%

\begin{abstract}
We propose the method for estimation of entropy generated during the
string breaking in high energy collisions. The approach is highly
based on the ideas proposed by Kharzeev D et al and may be useful in
thermalization problem.
\end{abstract}

\keywords{Ентропія зчепленості, термалізація, ефект
Унру/Entanglement entropy, thermalization, Unruh effect}

\maketitle

\section{Introduction}
Quark-gluon plasma, once being born in heavy ion collisions, quickly
(by the time of order of 1 fm) forgets about its initial conditions
and results in a completely thermal state. Loss of any information
but except of some total quantities (such as energy, angular
momentum, charge etc.) implies that the system meets with generation
of large amount of entropy right after the collision. To date there
is no complete theoretical explanation of this process known as
`thermalization'.

In \cite{t_ini} authors proposed the description of thermalization
which is based on the Unruh \cite{u_Unruh} effect. They proposed to
consider Unruh temperature $T_U$
\begin{equation}\label{T_U}
T_U=\frac{a}{2\pi}
\end{equation}
as the Hagedorn temperature $T_H$ (all over the paper we use planck
units); here $a$ is the acceleration caused with the momentum
exchanges during the collision.

The idea has been further developed in \cite{t_c-hor,t_c-hor_ext}.
Authors suggested that during the collisions string breaking leads
to parton deceleration $a$ and, consequently, to appearance of the
horizon. The deceleration can be expressed in a simple form via the
string tension $\sigma$. Due to the Unruh effect, see \eqref{T_U},
one meets with particle emission at the horizon. The partons have no
access to the region under the horizon thus meeting with the
information loss and, consequently, thermalization. The color
horizon (CH) radius, which was named as color confinement radius,
turned out to be of 1 fm order of magnitude. Comparison of the model
predictions to experimental data one can find in \cite{t_exp}.

However, the presence of horizon implies the existence of some kind
of black hole. Theoretical analysis of this idea and search for the
corresponding black hole type was made in \cite{t_str}.

Further development of the Unruh thermalization mechanism one can
find in \cite{t_cp_vio} where author considers P and CP violation
via Unruh mechanism. In \cite{t_gamma_quest} the discrepancy between
the approach and the RHIC data on photon radiation was revealed.

One may conclude that the Unruh thermalization is a promising
approach. However it can not explain completely all the data and
therefore needs further research and verification.

Here we utilize the analogy, proposed in the papers mentioned above,
between the CH generated in high energy collisions and the event
horizon (EH) of Schwarzschild black hole (BH). Our idea is highly
related to \cite{a_bh_entr_en,a_bh_entr_en_small} where scalar
radiation entropy from EH was considered from the Schmidt
decomposition viewpoint. Here we demonstrate how BH entropy can be
applied for the analysis of spinless radiation from CH -- as it is
``seen'' by the particles themselves. The approach might shed more
light on the thermalization problem.

The paper is organized as follows. In section 2 we present the
basics of the model. Section 3 is devoted to the asymptotics of the
scalar radiation entropy at CH. Conclusions and open questions one
can find in section 4.

\section{Formalism}
The approach is based on the model presented in
\cite{a_bh_entr_en,a_bh_entr_en_small}. In this section we utilize
the key formulae and concepts from the papers.

Let consider what happens during high energy collision;
consideration is restricted to spherically symmetric case for
simplicity.

Right after the collision the partons possess large momenta, thus
leading to the string breaking and to the generation of new
particles. The generation requires energy and therefore leads to the
loss of momenta with the initial partons and to their deceleration.
In \cite{t_c-hor,t_c-hor_ext} authors proposed to interpret such a
string breaking in the framework of Unruh effect and resulted in the
CH appearance; the idea has been developed further in
\cite{,t_exp,t_cp_vio,t_gamma_quest}.

Due to the Unruh effect, any acceleration is equivalent to the
presence of some EH and therefore implies the entropy generation.

To proceed one should analyze the interconnections between the
acceleration and event horizon in more details.

Let consider Schwarzschild BH with mass $M$. As it follows from
\cite{u_Unruh}, vacuum is not invariant with respect to different
frames of reference. The annihilation and creation boson operators
for free-falling observer $b$ and $b^\dag$ can be expressed with the
help of Bogoljubov transformations as
\begin{equation}\begin{split}\label{bb+}
&b=\frac{1}{\sqrt{1-\zeta^2}}c_{\rm
out}-\frac{\zeta}{\sqrt{1-\zeta^2}}\,c^\dag_{\rm in}\\
&b^\dag=\frac{1}{\sqrt{1-\zeta^2}}\,c^\dag_{\rm
out}-\frac{\zeta}{\sqrt{1-\zeta^2}}\,c_{\rm in}\end{split},
\end{equation}
where $c,\,c^\dag$ are the annihilation and creation boson operators
in the accelerated frame of reference, subscripts $_{\rm in(out)}$
determine the region in(out)side the BH horizon, and $\zeta$ is
defined as
\begin{equation}\label{zeta}
\zeta={\rm exp}\left(-4\pi M\omega\right),
\end{equation}
where $\omega$ is the energy of the quanta generated at the horizon.

Using \eqref{bb+} one can rewrite the vacuum state $|0\rangle\colon
b|0\rangle=0$ from the free-falling frame of reference as
\[
|0\rangle=\sqrt{\frac{1-\zeta^2}{1-\zeta^{2N}}}\sum_{n=0}^{N-1}\zeta^n|n\rangle_{\rm
in}|n\rangle_{\rm out},
\]
where $N$ stands for the dimension of the in(out)side Hilbert
subspace. As one can notice, this is just the Schmidt decomposition
of the vacuum state. The subscript $_{\rm in}$ denotes the inside
degrees of freedom which are located under the EH and thus are
inaccessible for the accelerated observer. Tracing out over them one
results in a mixture state described with density matrix $\rho_{\rm
out}$
\[
\rho_{\rm out}={\rm Tr}_{\rm
in}|0\rangle\langle0|=\frac{1-\zeta^2}{1-\zeta^{2N}}\sum_{n=0}^{N-1}\zeta^{2n}|n\rangle_{\rm
out}\langle n|
\]
with von Neumann entropy $\sigma\left(N,\zeta\right)$
\begin{equation}\label{sigma}
\sigma\left(N,\zeta\right)=-\ln\frac{1-\zeta^2}{1-\zeta^{2N}}
-\left(\frac{\zeta^2}{1-\zeta^2}-N\frac{\zeta^{2N}}{1-\zeta^{2N}}\right)\ln\zeta^2.
\end{equation}

Expression \eqref{sigma} defines the entropy for some mode with
angular momentum and energy $\omega$ being fixed, and therefore we
should sum up the contributions of all the modes available.

The angular momentum of the radiated quanta is restricted with
\[
0\leq\sqrt{l(l+1)}\leq\sqrt{L(L+1)}=rp=2M\sqrt{\omega^2-m^2},
\]
where $m$ is the rest mass of the particles radiated away from the
horizon. Summing over the angular degrees of freedom gives then
\[\begin{split}
&\sum_{l=0}^{l=L}\sum_{\mu=-l}^{\mu=l}1=
4M^2\left(\omega^2-m^2\right)+\\
&+\frac{\sqrt{16M^2\left(\omega^2-m^2\right)+1}+1}{2}.
\end{split}\]
Integrating over all the $\omega$ possible, finally we obtain the
scalar entropy $S\left(N,M,m\right)$
\begin{equation}\begin{split}\label{S}
&S\left(N,M,m\right)=\frac{M^2}{6\pi^3}\int_{\zeta_m}^{\zeta_M}{\rm
d}\zeta\frac{\sigma\left(N,\zeta\right)}{\zeta}\Biggl(\frac{\ln^2\zeta}{2\pi^2}+1-\\
&-8M^2m^2+\sqrt{\frac{\ln^2\zeta}{\pi^2}+1-16M^2m^2}\,\Biggr),
\end{split}\end{equation}
where $\zeta_m={\rm exp}(-4\pi Mm)$, $\zeta_M={\rm exp}(-4\pi M^2)$,
and $\zeta$ is defined in \eqref{zeta}.

Entropy $S\left(N,M,m\right)$ strongly depends on its arguments. The
integrand in \eqref{S} is highly correlated to the boundaries of
integration, and it is not easy to calculate $S\left(N,M,m\right)$
even numerically. In order to estimate the entropy and to compare it
to the experimental data from the collision experiments one should
know its dependence on the asymptotic values of the arguments.

\section{Asymptotic analysis}
In this section we present the asymptotics of \eqref{S}. Those who
are interested in details we refer to
\cite{a_bh_entr_en,a_bh_entr_en_small}.

Before we proceed let us make some general restrictions.

As it follows from \cite{t_c-hor,t_c-hor_ext}, Unruh temperature
$T_U$ can be expressed as $T_U=\sqrt{\sigma/2\pi}$, where $\sigma$
is the string tension. Using \eqref{T_U} one obtains then
\begin{equation}\label{M}
M=\frac{1}{4\sqrt{2\pi\sigma}},
\end{equation}
where we have used that for BH $a=1/(4M)$.

For the mass $m$ of the particles generated at the CH we have
\begin{equation}\label{m}
10^{-23}\leq m\leq10^{-17},
\end{equation}
where $10^{-23}$ stands for the electron mass, and $10^{-17}$ stands
for the mass of the Higgs boson.

One should keep in mind that masses $M$ and $m$ from \eqref{M} and
\eqref{m} are not independent: since BH can not emit the particle
with mass larger than its own then from \eqref{M}
\begin{equation}\label{m<M}
m\leq M\quad\Rightarrow\quad 4\sqrt{2\pi\sigma}\,m\leq1.
\end{equation}

Entropy \eqref{S} is valid for the scalar radiation, i.e. for the
spinless particles.

\subsection{Finite $N$}
In case of small BH, i.e. $M\leq1$, the amount of degrees of freedom
involved in the process will be small too. As a result $N$, i.e. the
dimension of the corresponding Hilbert space, will be not very
large.

Asymptotic expression for entropy $S\left(N,M,m\right)$ in case of
finite $N$ was obtained in \cite{a_bh_entr_en_small}. The
corresponding asymptotics was assumed to be valid for the
(sub)planck masses, i.e. for $M\leq 1$ (rigorous analysis reveals
that this limitation on $M$ is inessential). The corresponding
entropy $S(N,M,m)$ can be expressed then as
\begin{eqnarray}\label{S N}
S\left(N,M,m\right)\approx-\frac{M^2}{6\pi^2}\left.\left(\sum_{n=1}^{N}\alpha_n-N\alpha_N\right)\right|_{\omega=m}^{\omega=M},
\end{eqnarray}
where
\[\begin{split}
&\alpha_n=\frac{\zeta^{2n}}{n}\Biggl[1+4M\omega+8M^2\left(2\omega^2-m^2\right)+\\
&+32M^3\omega\left(\omega^2-m^2\right)+\\
&+\frac{1+6M\omega+8M^2\left(2\omega^2-m^2\right)}{\pi n}+\\
&+\frac{3+16M\omega}{4\pi^2n^2}+\frac{1}{2\pi^3n^3}\Biggr]+\left(1-8M^2m^2\right)\times\\
&\times\left(2\pi-1/n\right)e^{2\pi n}{\rm Ei}\left[-2\pi
n\left(1+4M\omega\right)\right],
\end{split}\]
where ${\rm Ei}(x)=\int_{-\infty}^xe^{-t}/t{\rm d}t$; $M$ is defined
from \eqref{M}.

To obtain \eqref{S N} the integrand in \eqref{S} was decomposed into
series. These series are convergent for any values of $N$, $M$ and
$m$ till \eqref{m<M} is valid and till $m\neq0$. But such weak
limitations may lead to large number of terms in the sum from
\eqref{S N}. It follows from the decomposition of expression
\eqref{sigma} into power series up to $\zeta^{2N}$. To provide quick
convergence of the series and to reduce the number of terms to be
taken into account we can use
\[
mN\sqrt{2\pi/\sigma}>1,
\]
where we used \eqref{M}. This inequality can be considered as a kind
of (weak) restriction for $N$.

In addition, as it follows from \cite{a_bh_entr_en_small}, the
following inequality should be valid:
\begin{equation}\label{cond Mm}
\frac{m^2\left(1+2m\right)^2}{32\pi\sigma\left(1+m/\sqrt{2\pi\sigma}\right)^4}\ll1,
\end{equation}
where we used \eqref{M}. This is necessary to simplify the square
root term in the integrand in \eqref{S} (see
\cite{a_bh_entr_en_small}, eq.(8) and the text right below it). As
it follows from \eqref{m<M}, \eqref{cond Mm} is valid by default;
however, it is necessary to provide good accuracy for the entropy
estimation.

\subsection{Infinite $N$}
In the collision experiments one usually results in large amount of
particles generated under the collision. It implies that the
dimension of the corresponding Hilbert subspace $N$ is large enough
to neglect the higher powers of $\zeta$ ($\zeta<1$ in case $m\neq0$,
see \eqref{zeta}) and thus usually we can put $N$ to be infinite. In
such a case from \eqref{sigma}
\begin{equation}\label{sigma inf}
\sigma\left(N\to\infty,\zeta\right)\approx-\ln\left(1-\zeta^2\right)-\frac{\zeta^2}{1-\zeta^2}\ln\zeta^2.
\end{equation}

Such an asymptotics requires the collision energy $\sqrt{s}$ to be
large, since otherwise the initial momenta of the partons will be
small to generate enough particles to provide the large dimension of
the Hilbert space.

In such a case we consider two possible cases: $Mm\ll1$ and
$Mm\gg1$.

For $Mm\ll1$ we use eq.(18) from \cite{a_bh_entr_en}:
\begin{equation}\begin{split}\label{Mm<<1}
&S\left(N\to\infty,\sigma,m\right)\approx1.825\times10^{-2}M^2\times\\
&\times\left(1-4.348M^2m^2\right)=\\
&=\frac{1.8\times10^{-4}}{\sigma}\left(1-0.043\frac{m^2}{\sigma}\right),
\end{split}\end{equation}
where we used \eqref{M} also.

For $Mm\gg1$ we can take eq.(22) from \cite{a_bh_entr_en}:
\[\begin{split}
&-\frac{4\left(2\pi-1\right)}{81\pi^3}M^4m^2e^{-8\pi Mm}\leq\\
&\leq
S\left(N\to\infty,M,m\right)\leq\frac{4}{81\pi^3}M^4m^2e^{-8\pi Mm}.
\end{split}\]
Due to the exponential boundaries the entropy in such an asymptotics
can be set to 0 with high accuracy, and therefore this case can be
neglected.

\section{Discussion and conclusions}
In the paper we determined the entanglement entropy generated at the
CH during high energy collisions. It is estimated in the frame of
reference of the interacting particles meeting deceleration due to
the string breaking mechanism. To date it is applicable for the
spinless radiation only.

The model utilizes spherical symmetry, so the analysis is restricted
with central ion or with particle-antiparticle collisions. The last
ones are preferable since the underlying formalism is based on the
Schwarzschild black hole, which have zeroth electrical charge. To
satisfy the electrical neutrality of CH one may consider
zeroth-charge particles as the outgoing radiation since otherwise we
will go outside the electrical neutrality. But such a restriction
seems to be irrational due to the experimental restrictions: usually
it is hard to collect data for neutral particles (take $\pi^0$ meson
for example).

In order to extend the amount of data available for the analysis one
can take into account the charged outgoing particles. In such a case
the CH neutrality is satisfied if one deals with large momenta
exchanges, since this implies a large amount of particles being
generated; then the total charge will be 0 on average.

The presented approach is an attempt of further development of the
ideas proposed by Satz H et al. The presented expressions allow
direct estimation of the entropy and therefore might be useful in
the thermalization problem. Surely, the need experimental
verification which is the topic of our further research.

\bibliography{d:/Max/PhD/Dissertation/Literature/bib}
\begin{flushright}
{\footnotesize Received 08.08.14}
\end{flushright}

\rezume{
М.В. Теслик, О.М. Теслик}{ЕНТРОПІЯ ТЕРМАЛІЗАЦІЇ ЧАСТИНОК І ЕФЕКТ
УНРУ} {Запропоновано метод оцінки ентропії, яка генерується під час
розриву струни у високоенергетичних зіткненнях. Підхід базується на
ідеях Харзєєва Д. та ін. і може бути застосований для розв'язку
проблеми термалізацї.}

\rezume{
М.В. Теслык, Е.Н. Теслык}{ЭНТРОПИЯ ТЕРМАЛИЗАЦИИ ЧАСТИЦ И ЭФФЕКТ
УНРУ} {Предложен метод оценки энтропии, генерируемой во время
разрыва струны в высокоэнергетических столкновениях. Метод основан
на идеях Харзеева Д и др. и может быть применён при решении проблемы
термализации.}
\end{document}